\documentclass[npg]{copernicus}
\usepackage{amsmath}
\usepackage{graphicx}
\usepackage{ifthen}
\usepackage{amsfonts}
\usepackage{amssymb}
\usepackage{epsfig}
\usepackage{url}
\usepackage{color}

\def\be{\begin{equation}}
\def\ee{\end{equation}}
\newtheorem{theorem}{Theorem}

\newtheorem{proposition}[theorem]{Proposition}

\begin{document}
\title{A delay differential model of ENSO variability, Part 2:\\
Phase locking, multiple solutions, and dynamics of extrema}
\author[1]{Ilya Zaliapin}
\author[2]{Michael Ghil}

\affil[1]{Department of Mathematics and Statistics,
University of Nevada, Reno, USA.
E-mail: zal@unr.edu.}

\affil[2]{Geosciences Department and Laboratoire de M\'{e}t\'{e}orologie 
Dynamique (CNRS and IPSL), Ecole Normale Sup\'{e}rieure, Paris, FRANCE, 
and Department of Atmospheric \& Oceanic Sciences and Institute of Geophysics 
\& Planetary Physics, University of California, Los Angeles, USA. 
E-mail: ghil@lmd.ens.fr.}

\runningtitle{A delay differential model of ENSO variability, Part 2}

\runningauthor{I. Zaliapin and M. Ghil}

\correspondence{Ilya Zaliapin (zal@unr.edu)}

\received{}
\pubdiscuss{} 
\revised{}
\accepted{}
\published{}

\firstpage{1}

\maketitle

\begin{abstract}
We consider a highly idealized model for
El Ni\~no/Southern Oscillation (ENSO) variability, as introduced
in an earlier paper.
The model is governed by a delay differential equation for
sea surface temperature $T$ in the Tropical Pacific, and it
combines two key mechanisms that participate in ENSO dynamics: 
delayed negative feedback and seasonal forcing.
We perform a theoretical and numerical study of the model 
in the three-dimensional space of its physically relevant parameters:
propagation period $\tau$ of oceanic waves across the Tropical Pacific, 
atmosphere-ocean coupling $\kappa$, and strength of seasonal forcing $b$.
Phase locking of model solutions to the periodic forcing is prevalent:  
the local maxima and minima of the solutions tend to occur at the 
same position within the seasonal cycle. Such phase locking is
a key feature of the observed El Ni\~no (warm) and La Ni\~na (cold) 
events.  
The phasing of the extrema within the seasonal cycle
depends sensitively on model parameters when forcing is weak.
We also study co-existence of multiple solutions for fixed model
parameters and describe the basins of attraction of the stable solutions
in a one-dimensional space of constant initial model histories.
\end{abstract}


\introduction[Introduction and motivation]
\label{Intro}
\subsection{Key ingredients of ENSO theory}
\label{ENSO_theory}
The El-Ni\~no/Southern-Oscillation (ENSO) phenomenon is the most prominent signal 
of seasonal-to-interannual climate variability. Its crucial role in climate dynamics and
its socio-economic importance were summarized in the first part of this study \citep{GZT08},
hereafter Part 1; see also \cite{Phil90,Glantz+91,Diaz92} and \cite{Cane05}, among others.

An international ten-year (1985--1994) Tropical-Ocean--Global-Atmosphere (TOGA) Program
greatly improved the observation \citep{MBC98}, theoretical modeling~\citep{Neel+94,Neel+98}, 
and prediction \citep{Latif94} of exceptionally strong El Ni\~no events. 
It has confirmed, in particular, that ENSO's significance extends far beyond the Tropical 
Pacific, where its causes lie. 

An important conclusion of this program was that --- in spite of  
the great complexity of the phenomenon and the differences between the spatio-temporal 
characteristics of any particular ENSO cycle and other cycles --- the state of the 
Tropical Pacific's ocean-atmosphere system could be characterized, mainly, by 
either one of two highly anticorrelated scalar indices. These two indices are a sea surface
temperature (SST) index and the Southern Oscillation Index (SOI): they capture 
the East--West seesaw in SSTs and that in sea level pressures, respectively; see, 
for instance, Fig. 1 of \cite{SG01}.

A typical version of the SST index is the so-called Ni\~no--3.4 index, which
summarizes the mean ``anomalies" ---  {\it i.e.}, the monthly-mean deviations 
from the climatological ``normal" --- of the spatially averaged SSTs over the region 
(170$\degree$W--120$\degree$W, 5$\degree$S--5$\degree$N) 
\citep{HT99,RS94,Tren97}. 

\begin{figure}[t]
\vspace*{2mm}
\centering\includegraphics[width=8.3cm]{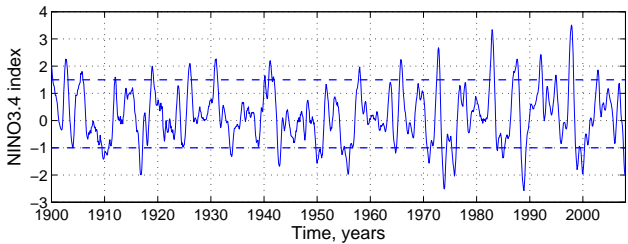} 
\caption{Temporal evolution of the NINO3.4 index that summarizes sea surface
temperature (SST) anomalies in the region between 
170$\degree$W--120$\degree$W and 5$\degree$S--5$\degree$N. The
time series is centered and normalized, but the horizontal lines do not 
represent the standard deviations: instead, they have have ordinates 1.5 and -1; 
see also Fig.~\ref{fig_nino_hist}.} 
\label{fig_nino}
\end{figure}

The evolution of this index since 1900 is shown in Fig.~\ref{fig_nino}: 
it clearly exhibits some degree of regularity, on the one hand, as well as numerous 
features characteristic of a deterministically chaotic system, on the other. The
regularity manifests itself as the rough superposition of two dominant oscillations
--- quasi-biennial and quasi-quadrennial \citep{Jiang+95,Ghil+02} --- accompanied
by a near-symmetry of the local maxima and minima ({\it i.e.}, of the positive and 
negative peaks). The lack of regularity has been associated with the presence
of a ``Devil's staircase" \citep{JNG94,JNG96,Tzi+94,Tzi+95} and does not preclude
the superposition of stochastic effects as well \citep{GCS08}.

While this study mainly focuses on local {\it extrema} (maxima and minima) 
in our ENSO model, one must recall that the major El Ni\~nos of 1982-83 and 1997-98 
(see Fig.~\ref{fig_nino}) are, in fact, genuine {\it extremes}, {\it i.e.} rare events of unusually 
large magnitude. 
These climatic extremes and the related hydroclimatological impacts are part of
the motivation for studying ENSO in general and for this study in particular.
At the moment, the observational record contains too few of these truly extreme 
events to allow studying them by the methods of classical, {\it i.e.} statistical extreme 
value theory. We hope, therefore that the modeling approach developed in this study 
might prove useful in obtaining relevant statistical data for better understanding 
ENSO-related extreme events.

To simulate, understand and predict such complex phenomena one needs 
a full hierarchy of models, from ``toy'' via intermediate to fully coupled general 
circulation models (GCMs)~\citep{Neel+98,GR00}.
We focus here on a ``toy'' model, which captures a qualitative, conceptual picture 
of ENSO dynamics that includes a surprisingly broad range of features. 
This approach allows one to gain a rather comprehensive understanding 
of the model's, and maybe the phenomenon's, underlying mechanisms and 
their interplay, at the cost of not capturing a full spatio-temporal picture of 
ENSO evolution.

We consider the following conceptual ingredients that play a determining 
role in the dynamics of the ENSO phenomenon: 
(i) the Bjerknes hypothesis, which suggests a positive feedback as a mechanism for 
the growth of an internal instability that could produce large positive anomalies 
of SSTs in the eastern Tropical Pacific \citep{Bjer69};
(ii) delayed oceanic wave adjustments, realized in the form of eastward Kelvin and 
westward Rossby waves, that compensate for Bjerknes's positive feedback \citep{SS88}; 
and (iii) seasonal forcing \citep{Batt88,Cha94,Cha95,JNG94,JNG96,Tzi+94,Tzi+95,GR00}. 
A more detailed discussion of these ingredients is given by \cite{GZT08} and references therein.

The past 30 years of research have shown that ENSO dynamics is governed, by and large, 
by the interplay of the above nonlinear mechanisms and that their simplest version can 
be studied in periodically forced Boolean delay systems \citep{SG01,GZC08} and delay differential 
equations (DDE) \citep{SS88,BH89,Tzi+94}. 
DDE models provide a convenient paradigm for explaining interannual ENSO 
variability and shed new light on its dynamical properties. 
So far, though, DDE model studies of ENSO have been limited to linear stability analysis 
of steady-state solutions, which are not typical in forced systems; case studies of 
particular trajectories; or one-dimensional (1-D) scenarios of transition to chaos, 
where one varies a single parameter while the others are kept fixed. 
A major obstacle for the complete bifurcation and sensitivity analysis of DDE models 
lies in the complex nature of DDEs, whose analytical and numerical treatment is  
considerably harder than that of their ordinary differential equation (ODE) counterparts.

\begin{figure*}[t]
\vspace*{2mm}
\centering\includegraphics[height=5cm]{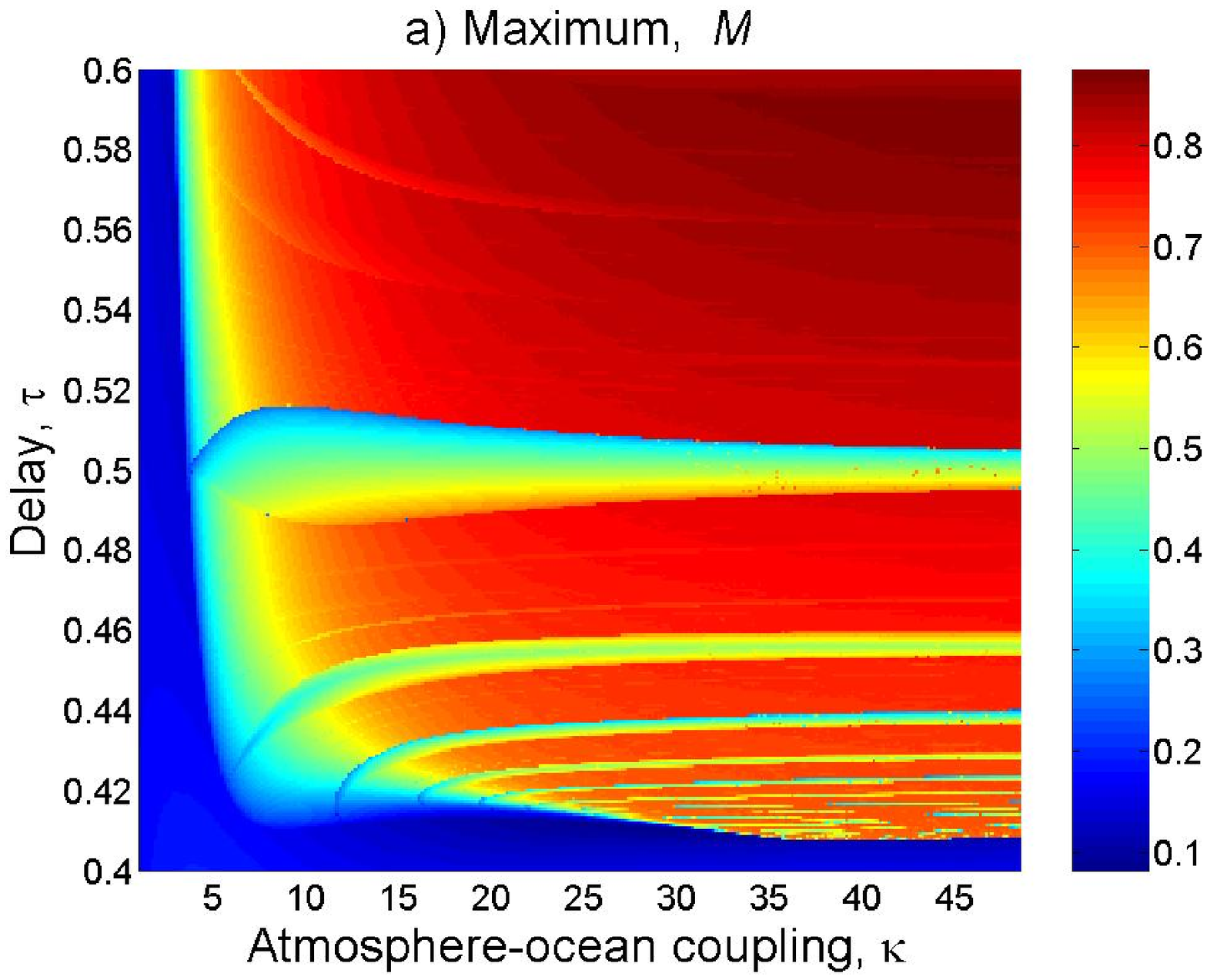}
\centering\includegraphics[height=5cm]{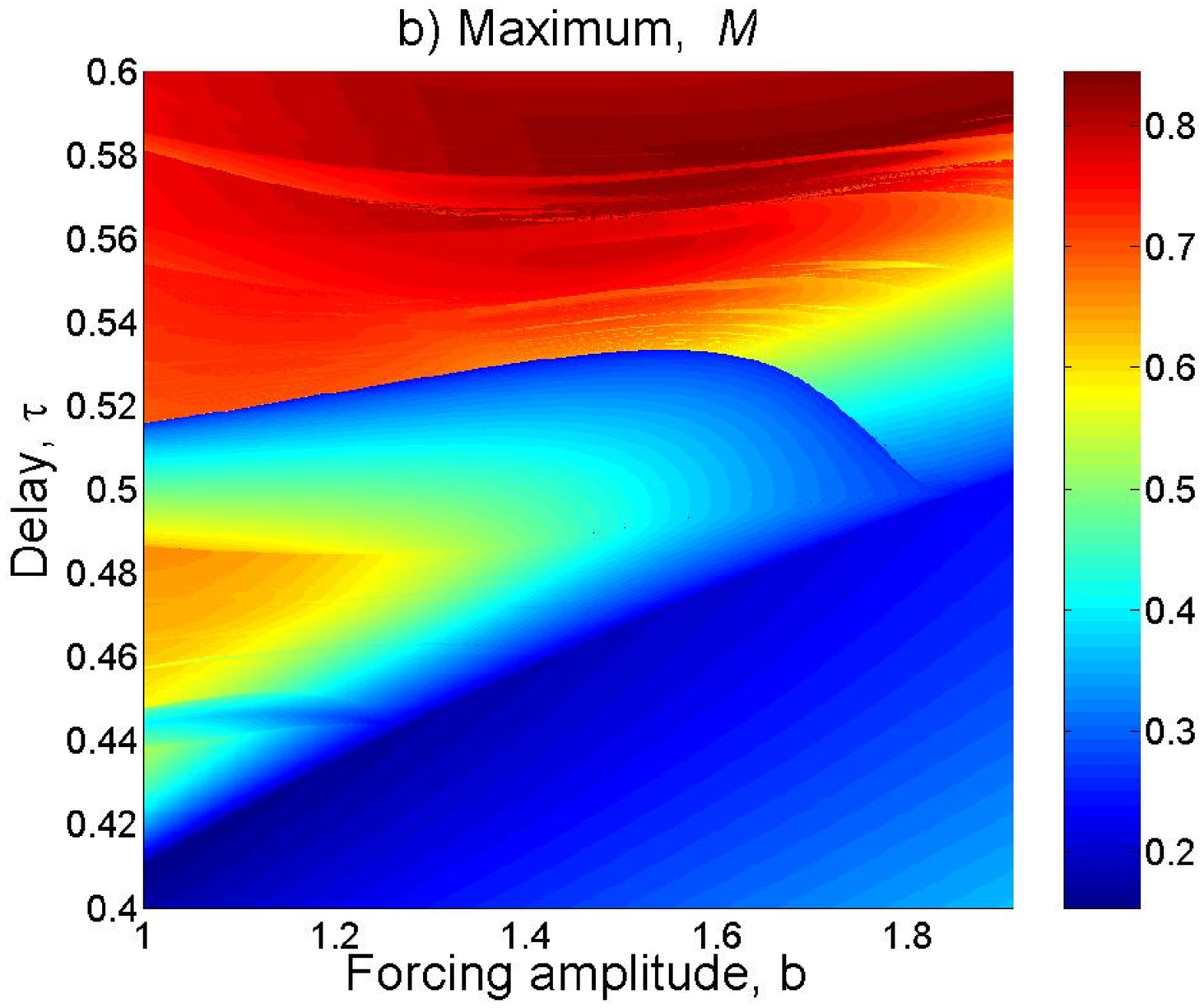}
\centering\includegraphics[height=5cm]{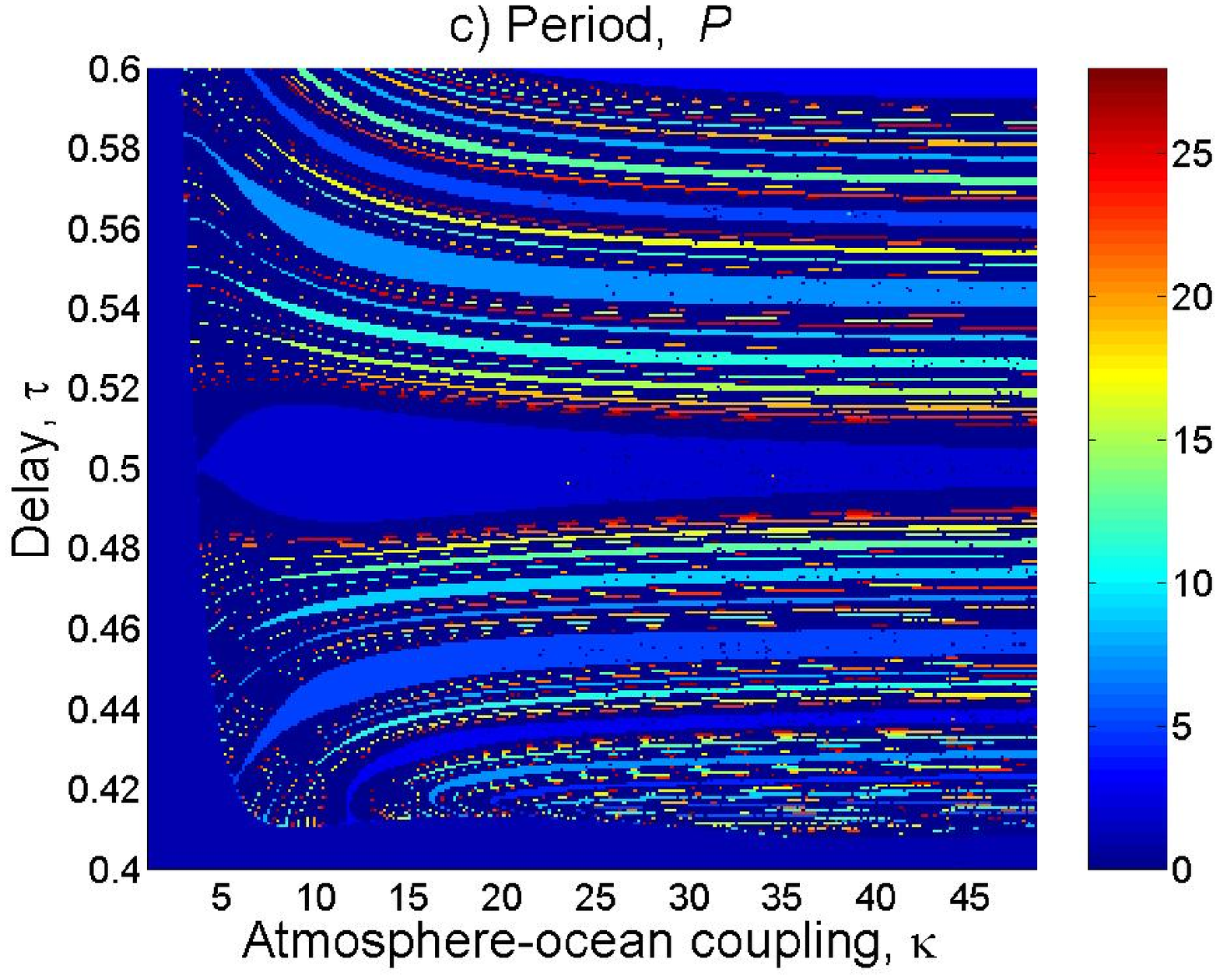}
\centering\includegraphics[height=5cm]{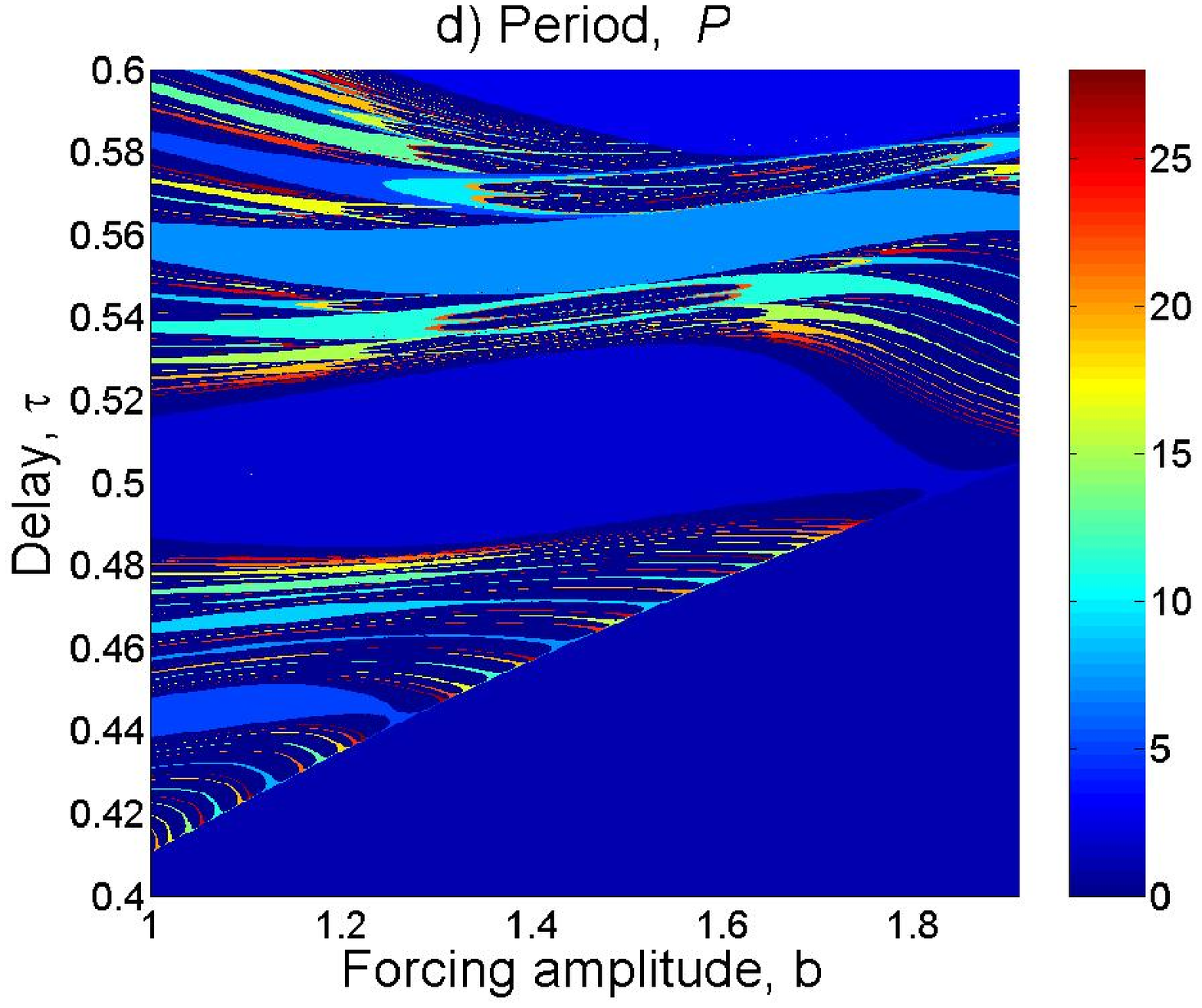}
\caption{Maximum and period maps for a warm initial history, $\phi(t)\equiv 1$.
($a$) Maximum map, $M=M(\kappa,\,\tau)$ at $b=1$;
($b$) Maximum map, $M=M(b,\,\tau)$ at $\kappa=10$;
($c$) Period map, $P=P(\kappa,\,\tau)$ at $b=1$;
($d$) Period map, $P=P(b,\,\tau)$ at $\kappa=10$. 
Reproduced from \cite{GZT08}, with kind permission of Copernicus Publications on behalf of 
the European Geosciences Union (EGU).} 
\label{stat_map_a}
\end{figure*}

\subsection{Part 1 results and their physical interpretation}
\label{Part_1}
\cite{GZT08} took several steps toward a comprehensive analysis,
numerical as well as theoretical, of a DDE model relevant for ENSO phenomenology. 
In doing so, they also illustrated the complexity of the structures that arise in its
phase-and-parameter space for even such a simple model of climate dynamics.
Specifically, the authors formulated a highly idealized DDE model for 
ENSO variability and focused on the analysis of model solutions in a broad 
three-dimensional (3-D) domain of its physically relevant parameters. 
They showed that this model can reproduce many scenarios relevant to
ENSO phenomenology, including prototypes of El Ni\~no and La Ni\~na
events; spontaneous interdecadal oscillations; and intraseasonal activity 
reminiscent of Madden-Julian oscillations or westerly wind bursts.

This model was also able to provide a good justification for the observed 
quasi-biennial oscillation in Tropical Pacific SSTs and trade winds 
\citep{Phil90,Diaz92,Jiang+95,Ghil+02}, with the 2--3-year period arising
naturally as the correct multiple of the sum of the basin transit times of Kelvin and Rossby waves.
An important finding of \cite{GZT08} was the existence
of regions of stable and unstable solution behavior in the model's parameter space;
these regions have a complex and possibly fractal distribution of solution properties.

Figure~\ref{stat_map_a} illustrates the model's sensitive dependence on 
parameters in a region that corresponds roughly to actual ENSO
dynamics. The figure shows the behavior of the global maximum $M$ and period $P$ of 
model solutions as a function of three parameters: the propagation period $\tau$ of 
oceanic waves across the Tropical Pacific, the atmosphere-ocean coupling strength 
$\kappa$, and the amplitude $b$ of the seasonal forcing; for aperiodic solutions we set $P=0$.
Although the model is sensitive to each of these three parameters, sharp variations
in $M$ and $P$ are mainly associated with changing the delay $\tau$, which is 
plotted on the ordinate in all four panels of the figure.
In other words, the global maximum, in panels (a) and (b), as well as the period,
in panels (c) and (d), may change more than twofold in response to a slight variation 
of $\tau$.

This sensitivity is an important qualitative conclusion since in reality the propagation times
of Rossby and Kelvin waves are affected by numerous phenomena that are not related
directly to ENSO dynamics. Moreover, the instabilities disappear and the dynamics of 
the system becomes purely periodic, with period one year, as soon as the atmosphere-ocean 
coupling $\kappa$ vanishes or the delay $\tau$ decreases below a critical value;
see Figs.~\ref{stat_map_a}a, b.
Finally, the boundary between the domains of stable and unstable model behavior
is clearly visible in Fig.~\ref{stat_map_a}, in the lower-right part of panels (b) and (d). 

The region below and to the right of this boundary contains simple period-one 
solutions that change smoothly with the values of model parameters. 
The region above and to the left is characterized by sensitive dependence on
parameters.
The range of parameters that corresponds to present-day ENSO dynamics 
lies on the border between the model's stable and unstable regions.
Hence, if the dynamical phenomena found in the model have any relation to
reality, Tropical Pacific SSTs and other fields that are highly correlated with them,
inside and outside the Tropics, can be expected to behave in an intrinsically 
unstable manner; they could, in particular, change quite drastically with global
warming.

There are basically two approaches to ENSO dynamics \citep{Neel+94,Neel+98},
both of which may be useful in extending the results of Part 1 above.
The model considered here and in \cite{GZT08} explains the complexities of ENSO 
dynamics by the interplay of two oscillators: an internal, highly nonlinear one, due to 
a delayed feedback, and a forced, seasonal one. Our model thus falls within the 
strongly nonlinear, deterministic approach. 

An alternative approach attempts to explain several featureas of ENSO dynamics by 
the action of fast, ``weather" noise on a linear or very weakly nonlinear ``slow"
system, composed mainly on the upper ocean near the equator. 
\citet{BML04} and \citet{L+04}, among others, provide a comprehensive discussion of  
how weather noise could be responsible for the complex dynamics of ENSO, and, 
in particular, whether wind bursts trigger El Ni\~no events.
\citet{SKM06} explore this possibility in a conceptual toy model. 
\citet{GR00} already discussed the arguments about a ``stochastic paradigm'' 
for ENSO, with linear or only mildly nonlinear dynamics being affected decisively 
by weather noise, {\it vs.} a ``deterministically chaotic paradigm,'' with decisively 
nonlinear dynamics. \citet{GCS08} have recently illustrated a way of combining 
these two paradigms to obtain richer and more complete insight into climate dynamics 
in general.

The present paper continues the study initiated in Part 1 
and focuses (i) on the multiplicity of model solutions for the same parameter values; 
and (ii) on the behavior of local extrema in these solutions.
In particular, we investigate the distribution in time of the model solutions' maxima 
and minima; these extrema are directly connected to the strength and timing of the 
corresponding El Ni\~no (warm) or La Ni\~na (cold) events.
The current analytic theory of DDEs does not allow one to easily 
answer many practically relevant questions about the behavior of 
even such apparently simple equations as our Eq.~\eqref{dde} below.
The present study combines, therefore, general theoretical results about the existence
and continuous dependence of solutions on parameters with extensive 
numerical investigations.  

The rest of the paper is organized as follows.
In Section \ref{model}, we summarize the model formulation from Part 1, recall 
basic theoretical results concerning this model's solutions, and briefly review  
details of the numerical integration method.
Section~\ref{lock} reports on the phase locking of solutions to the periodic forcing,
namely on the tendency for the solutions' maxima and minima to each occur 
within a fixed, small interval of the seasonal cycle.
Existence of multiple solutions and the attractor basins of the stable solutions
are studied in Sect.~\ref{multi}.
In Sect.~\ref{discrete} we investigate the behavior of the model's local extrema, 
considered as a discrete dynamical system. A discussion of these results in
Sect.~\ref{discussion} concludes the paper.

\begin{figure}[t]
\vspace*{2mm}
\centering\includegraphics[width=8.3cm]{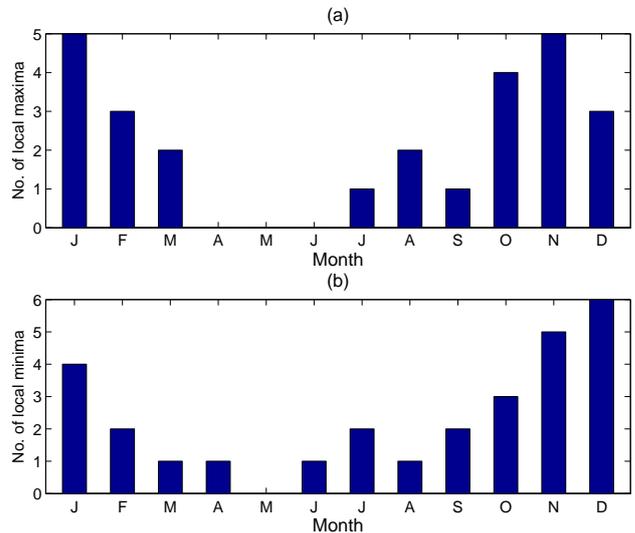} 
\caption{Histogram of temporal location of ($a$) warm and ($b$) cold events 
for the Ni\~no--3.4 index. 
The event thresholds are shown by the dashed horizontal lines in Fig.~\ref{fig_nino}.
Notice the preferential occurrence of both warm and cold events during the 
boreal winter.} 
\label{fig_nino_hist}
\end{figure}

\section{Model and numerical integration method}
\label{model}

\subsection{Model formulation and parameters}
Following Part 1, we consider a nonlinear DDE with additive, periodic forcing,
\begin{equation}
h'(t)=-a\,\tanh\left[\kappa\,h(t-\tau)\right]+b\,\cos(2\pi\,\omega\,t),
\label{dde}
\end{equation}
where $h'(t)={{\mathrm d}h(t)}/{{\mathrm d}t}$, $t\ge 0$, and the parameters 
$a,\tau,\kappa,b,$ and $\omega$ are all real and positive.
Equation~\eqref{dde} is a simplified one-delay version of 
the two-delay model considered by Tziperman {\it et al.} (\citeyear{Tzi+94});
it includes two mechanisms essential for ENSO variability: 
a delayed, negative feedback via the function $\tanh(\kappa\,z)$, and periodic external forcing.
As shown in Part 1, these two mechanisms suffice to generate very rich behavior
that includes several important features of more detailed models and of
observational data sets.

The function $h(t)$ in \eqref{dde} represents the thermocline depth deviations 
from the annual mean in the eastern Tropical Pacific; accordingly, it can also be 
interpreted roughly as the regional SST, since a deeper thermocline 
corresponds to less upwelling of cold waters, and hence higher SST, and
vice versa. The thermocline depth is affected by the wind-forced, eastward 
Kelvin and westward Rossby oceanic waves. 
The waves' delayed effects are modeled by the function 
$\tanh\left[\kappa\,h(t-\tau)\right]$;
the delay $\tau$ is due to the finite velocity of these waves and it corresponds 
roughly to their combined basin-transit time. 

The particular form of the delayed nonlinearity plays an important role in 
the behavior of a DDE model. \cite{MCZ91} provided a physical justification
for the monotone, sigmoid nonlinearity we adopt here.
The parameter $\kappa$, which is the linear slope of $\tanh(\kappa\,z)$
at the origin, reflects the strength of the atmosphere-ocean coupling.
The forcing term represents the seasonal cycle in the trade winds, with 
the strongest winds occurring in boreal fall. 

The DDE model \eqref{dde} is fully determined by its five parameters: 
feedback delay $\tau$, atmosphere-ocean coupling strength  $\kappa$, 
feedback amplitude $a$, forcing frequency $\omega$, and 
forcing amplitude $b$. 
By an appropriate rescaling of time $t$ and dependent variable $h$, we let 
$\omega=1$ and $a=1$. 
The remaining three parameters --- $\tau$, $\kappa$, 
and $b$ --- may vary, reflecting different physical conditions of 
ENSO evolution. We consider here the same parameter ranges as in Part 1
of this study:
$0\le \tau \le 2$~yr, 
$0< \kappa <\infty$,
$0\le b < \infty$.

To completely specify model \eqref{dde} we need to prescribe
some initial ``history,'' {\it i.e.} the behavior of $h(t)$ on the 
interval $[-\tau,\,0)$, {\it cf.} \cite{Hale}.
In the numerical experiments of Sect. \ref{lock} below we assume,
as in Part 1, that $h(t)\equiv 1$, 
$-\tau\le t<0$, {\it i.e.} we start with a warm year. But in Sect. \ref{multi}
we turn to a systematic exploration of the effect of the initial histories on
the number and stability of solutions.

\begin{figure}[t]
\centering\includegraphics[width=8.3cm]{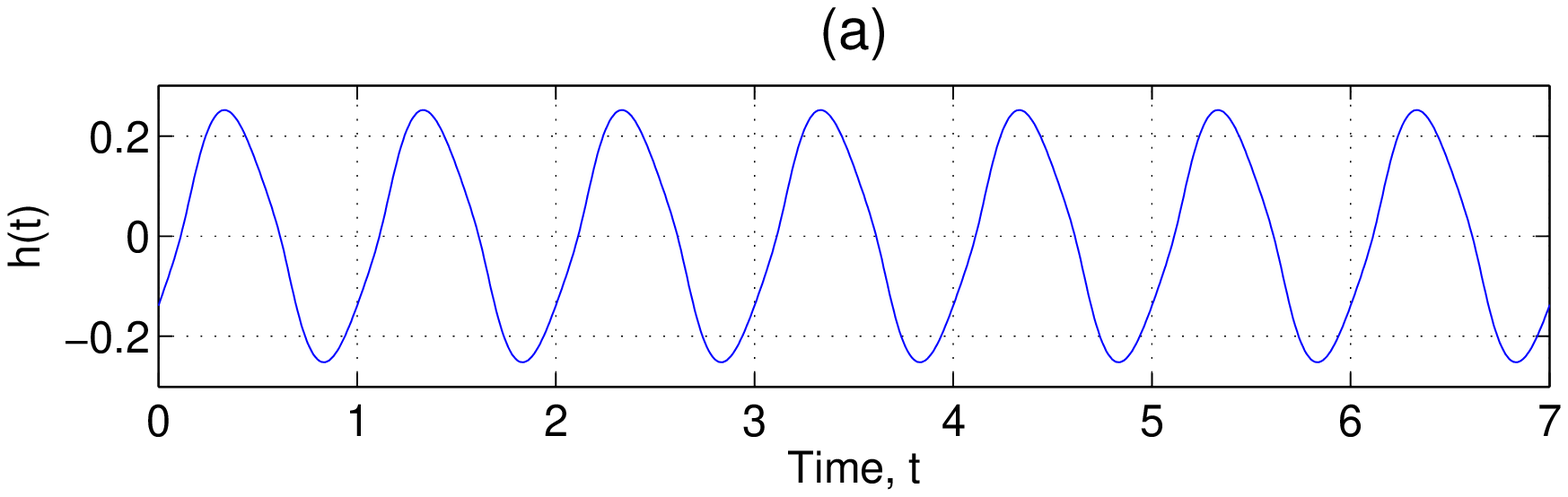} 
\centering\includegraphics[width=8.3cm]{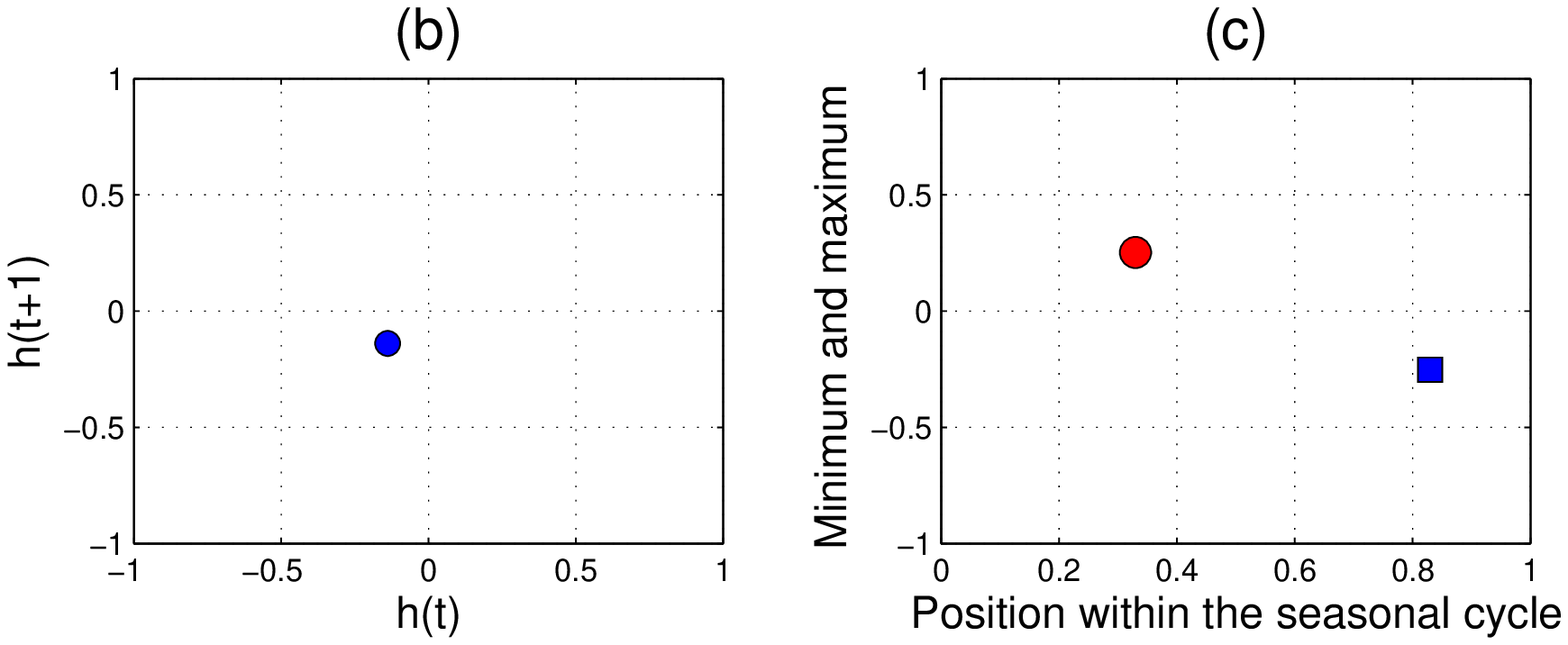}
\centering\includegraphics[width=8.3cm]{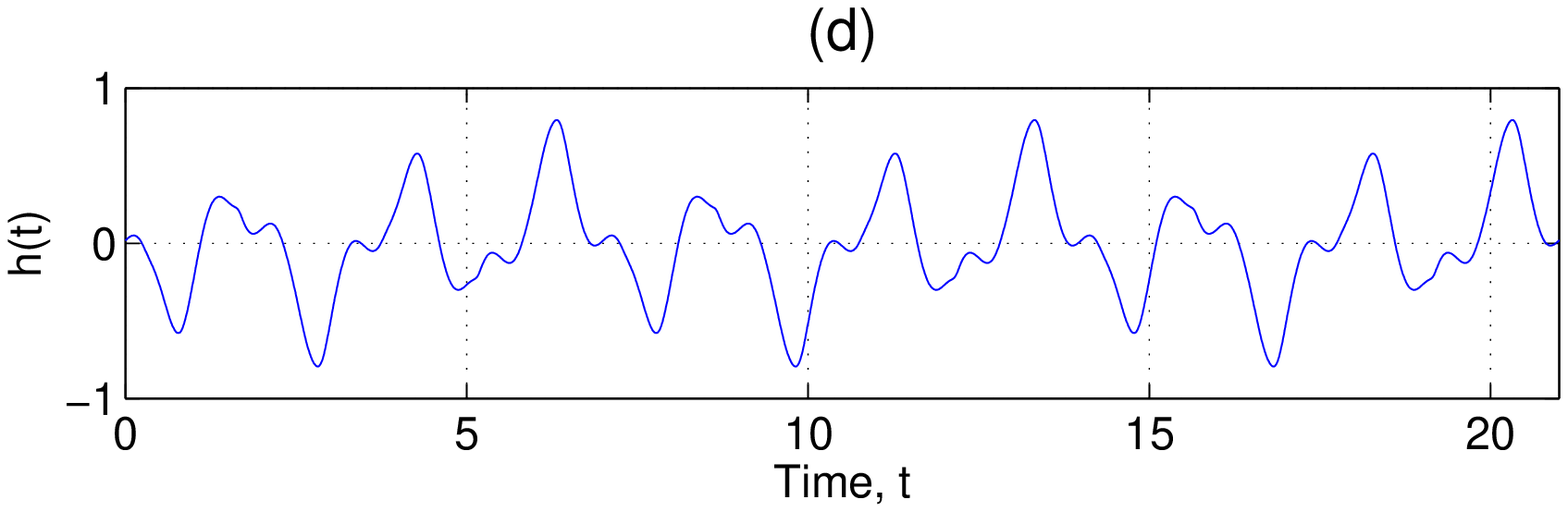}
\centering\includegraphics[width=8.3cm]{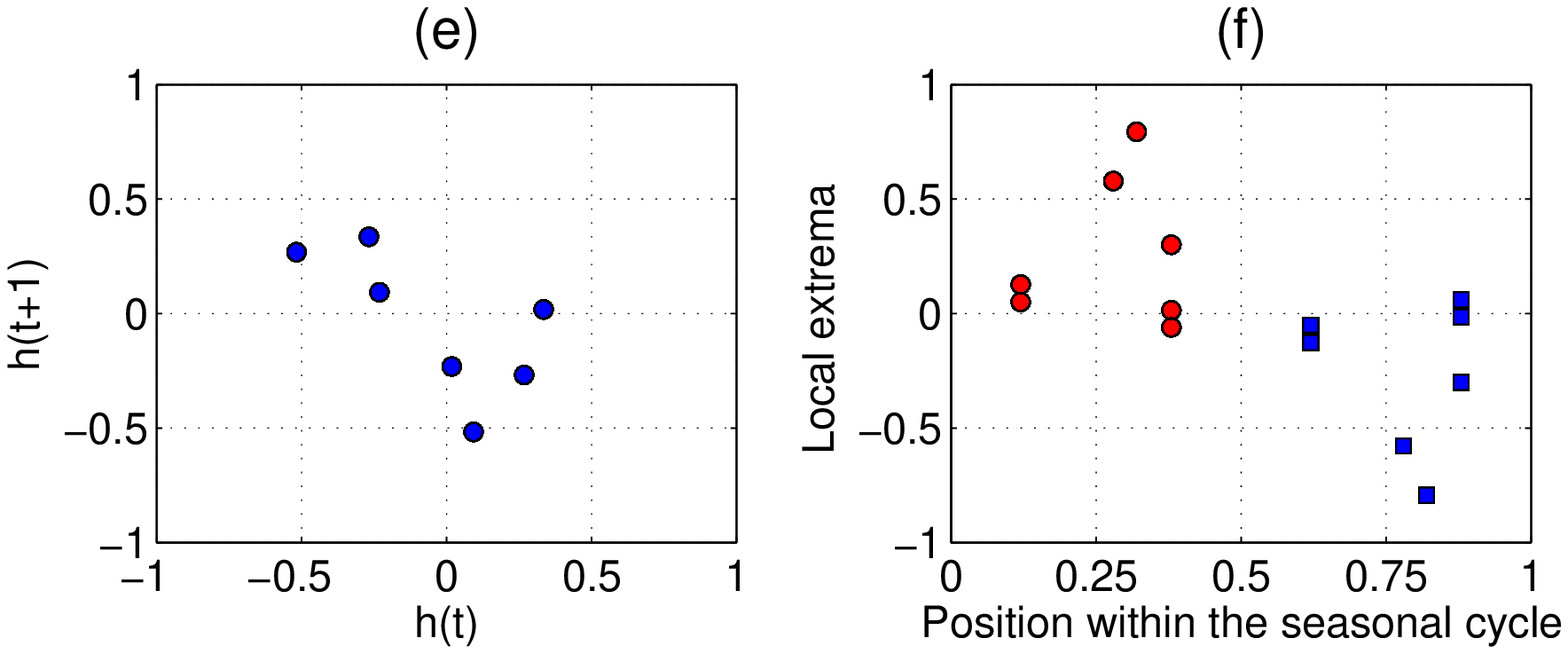}
\centering\includegraphics[width=8.3cm]{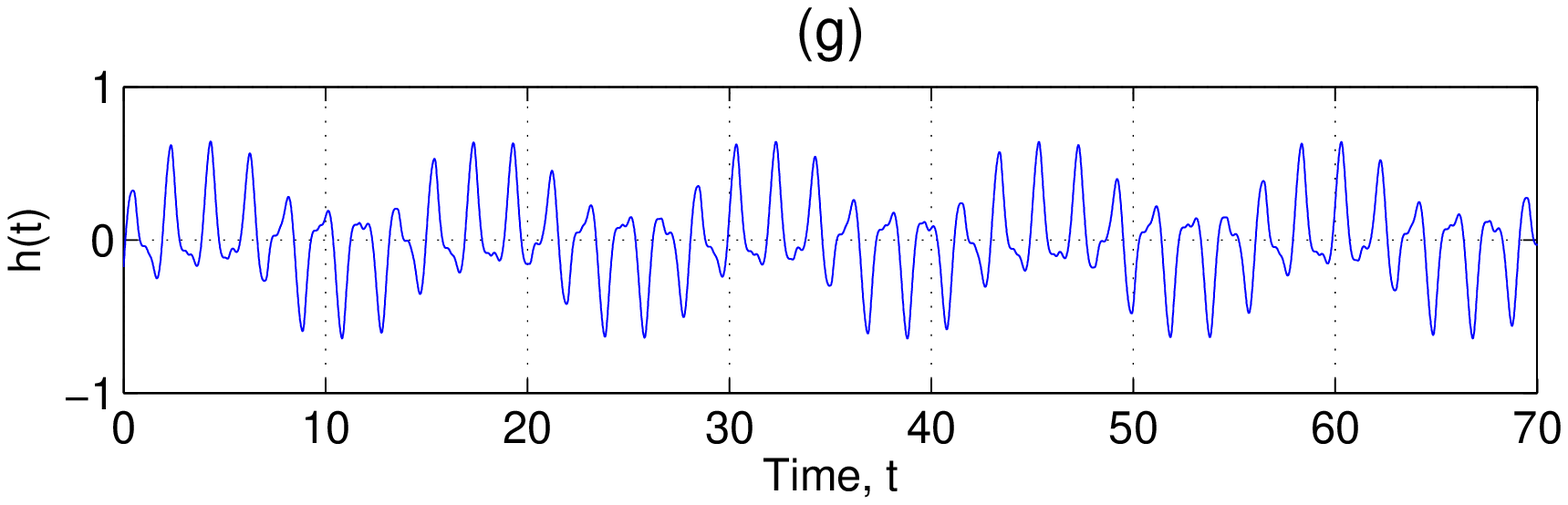}
\centering\includegraphics[width=8.3cm]{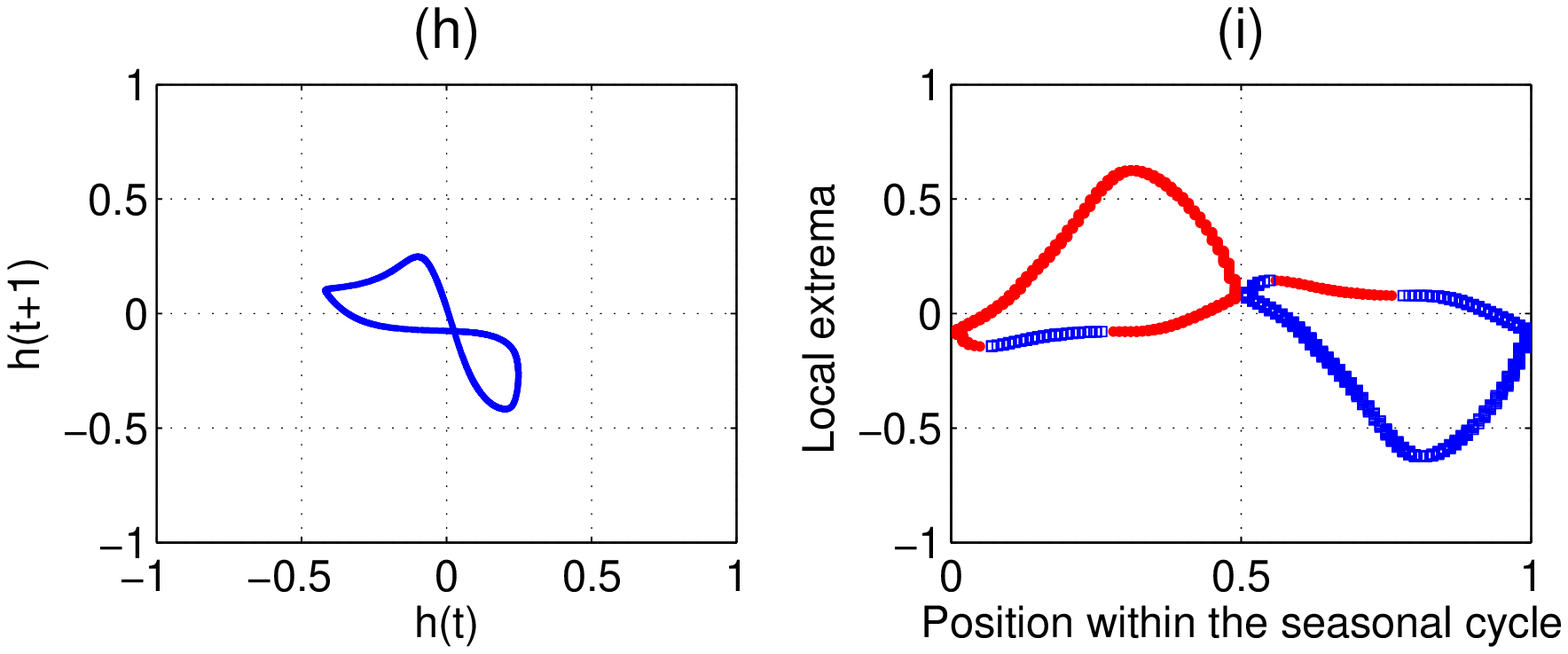}
\end{figure}

\begin{figure}[h]
\vspace*{2mm}
\caption{Seasonal phase locking of local extrema for model trajectories:
({\it a--c}) with period $P=1$;
({\it d--f})  with period $P=7$; and ({\it g--i}) aperiodic.
The model solutions in panels ($a, d, g$) are shown in the stationary regime, 
after a sufficiently long transient, and the time axis is shifted so as to 
start from $t=0$. The parameter values for these solutions are
($a$) $\tau=0.5$, $\kappa=11$, $b=2$; 
($d$) $\tau=0.56$, $\kappa=11$, $b=1.4$; and
($g$) $\tau=0.47$, $\kappa=10$, $b=1.0$.
The scatterplots of the points $(h(t_i), h(t_i+1))$ in panels ($b, e, h$) use the values
$i=0,1,\dots,500$, which correspond to $t_0=2500$ and the parameter
settings in panels ($a, d, g$), respectively. The phase locking is illustrated in
panels ($c, f, i$), which give the $h$-value of the local extrema --- maxima 
shown as red filled circles and minima as blue squares --- as a 
function of their position within the seasonal cycle, $\varphi=t({\rm mod}~1)$.} 
\label{tr_ex1}
\end{figure}

\subsection{Main theoretical result}
\label{Banach}
Consider the Banach space $X=C([-\tau,0),\mathbb{R})$ of continuous functions 
$h\,:\,[-\tau,0)\to\mathbb{R}$ and define
the norm for $h\in X$ as
\[\parallel h\parallel=\sup\left\{|h(t)|,~t\in[-\tau,0)\right\},\]
where $|\cdot|$ denotes the absolute value in $\mathbb{R}$ \citep{Hale,Nuss}.
For convenience, we reformulate the DDE initial-value problem (IVP)
in its rescaled form:
\begin{eqnarray}
h'(t)&=&-\tanh\left[\kappa\,h(t-\tau)\right]+b\,\cos(2\pi\,t),~t\ge 0,\label{ivp1}\\
\quad h(t)&=&\phi(t)~~{\rm for~}~t\in[-\tau,\,0),\quad\phi(t)\in X\label{ivp2}.
\end{eqnarray}

Ghil {\it et al.} (\citeyear{GZT08}) proved the following result, which follows 
from \cite{HaleVerd} and references therein.

\begin{proposition}{\bf (Existence, uniqueness, continuous dependence)}
For any fixed positive triplet $(\tau,\,\kappa,\,b)$, 
the IVP \eqref{ivp1}-\eqref{ivp2} has a unique solution $h(t)$ 
on $[0,\,\infty)$.
This solution depends continuously on the initial data $\phi(t)$, 
delay $\tau$ and the right-hand side of \eqref{ivp1}, considered as a continuous map
$f\,:\,[0,T)\times X\to \mathbb{R}$, for any finite $T$.
\label{prop}
\end{proposition}

From Proposition~\ref{prop} it follows, in particular, that the 
system \eqref{ivp1}-\eqref{ivp2} has a unique solution for all time, 
which depends continuously on the model parameters $(\tau,\,\kappa,\,b)$
for any finite time.
This result implies that any discontinuity in the solution profile, as
a function of the model parameters, indicates the existence of an
unstable solution that separates the attractor basins of two stable solutions.
Our numerical experiments suggest, furthermore, that all stable solutions 
of \eqref{ivp1}-\eqref{ivp2} are bounded and have an infinite number of zeros.

\subsection{Numerical integration}
\label{num}
The results in this Part 2 of our study are based on numerical integration of the 
DDE \eqref{ivp1}-\eqref{ivp2}.
We emphasize that there are important differences between the numerical
integration of DDEs and ODEs, and that these differences require developing
special software; often the problem-specific modification of such software
also becomes necessary.
We used here the Fortran 90/95 DDE solver \texttt{dde\_solver} of 
Shampine and Thompson (\citeyear{F90}), available at
\url{http://www.radford.edu/~thompson/ffddes/}.
Technical details of \texttt{dde\_solver}, as well as a brief overview 
of other available DDE solvers, are given in Appendix C of Part 1.

\begin{figure}[t]
\vspace*{2mm}
\centering\includegraphics[width=8.3cm]{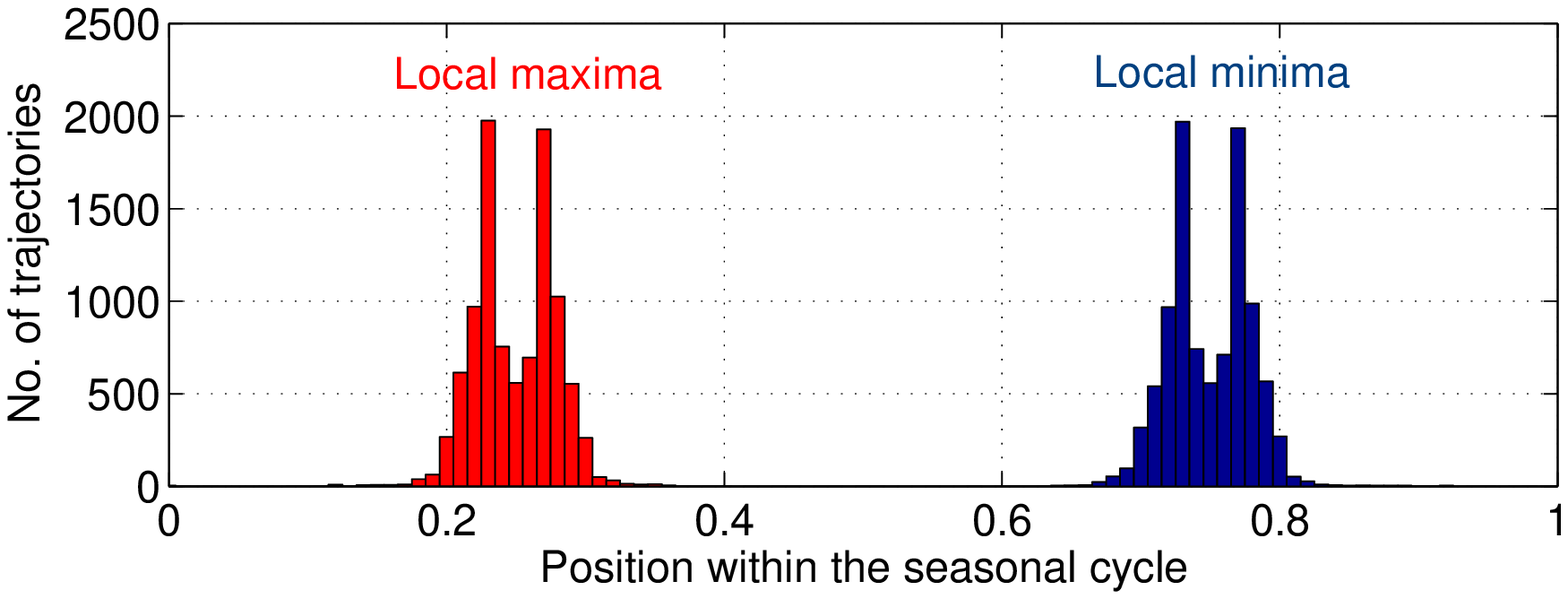}
\centering\includegraphics[width=8.3cm]{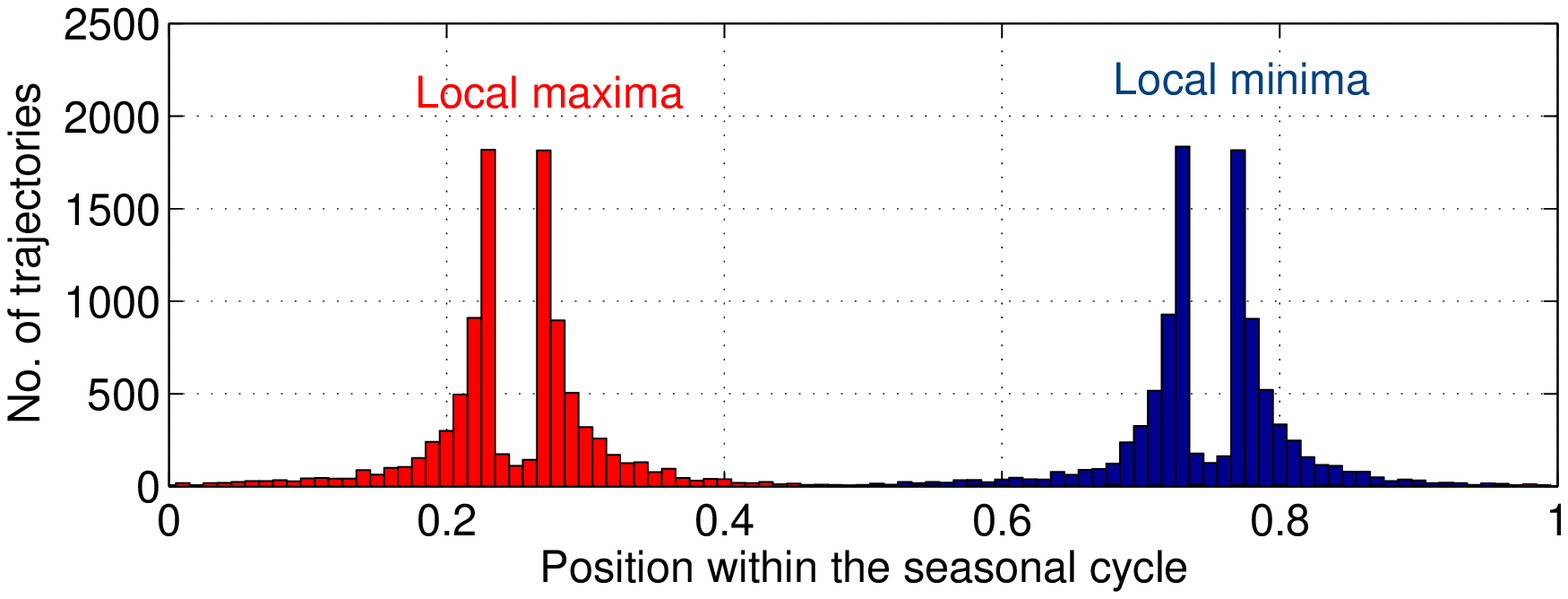}
\caption{Seasonal phase locking of local extrema: cumulative results.
Histogram of the phase $\varphi$ of the local maxima (red bars) and
minima (blue bars) of model solutions 
with $\kappa=2.0$ (top panel) and $\kappa=11.0$ (bottom panel).
Each panel uses 10 000 individual solutions with parameters 
$0<\tau\le 2$ and $0< b \le 10$; see also Fig.~\ref{expos_map}.} 
\label{expos_hist}
\end{figure}

\section{Seasonal phase locking of extrema}
\label{lock}
A distinctive feature of the extreme ENSO phases --- {\it i.e.}, of the El Ni\~no 
and La Ni\~na events --- is their occurrence during a boreal winter.
This phenomenon is illustrated in Fig.~\ref{fig_nino_hist}, which shows
the histograms of the monthly positions of unusually warm and unusually
cold events, based on the Ni\~no--3.4 index of Fig.~\ref{fig_nino}. In our 
classification, El Ni\~nos (see panel $a$) are those for which NINO3.4 $>1.5$,
while La Ni\~nas (see panel $b$) have NINO3.4 $<-1$.
This asymmetry in the classification is 
due to the fact that extreme warm events are more intense but fewer in number 
than the extreme cold events \citep{HKZ97,BS99,SCP00,KKGR05}.
Clearly, the extreme events, both warm and cold, tend to occur during 
boreal winter.

In discussing extrema, we distinguish between local and global ones. 
Recall that for a function $h(t)$ specified on the interval 
$[t_1,t_2]$, its {\it global} maximum (minimum) is defined as the point $t$ 
such that $h(t)$ is above (below) all the other values on that interval:
$h(t) \ge h(s)$, respectively $h(t)\le h(s)$, for all $s\in[t_1,t_2]$.
A {\it local} maximum (minimum) is a point $t$ such that the corresponding
value $h(t)$ is above (below) all the values in a vicinity of $t$;
for a sufficiently smooth function, the latter definitions are equivalent to  
\[{\rm (i)} \quad h'(t) = 0; \quad {\rm and} \quad
{\rm (ii)} \quad h''(t) < 0 \quad {\rm or} \quad h''(t)>0,\]
where $h''=(h')'$ is the second derivative of $h(t)$.

In this paper, we work with numerical solutions of the DDE problem \eqref{ivp1}-\eqref{ivp2}
that are available only on a finite time interval $[0,t_f]$;
in addition, we eliminate the initial transient interval $[0, t_0)$.
We thus consider the global and local extrema of our solutions
only on the interval $[t_0, t_f]$. 
The global extrema thus defined might differ in certain cases from their 
counterparts on the interval $[0,\infty)$, for which our DDE is formally defined.
The difference will only be noticeable for very--long-periodic, highly fluctuating 
solutions that are relatively rare in our model. Hence,
the reduced definitions of the global and local extrema do not affect the main 
conclusions of our analysis.

\begin{figure}[t]
\vspace*{2mm}
\centering\includegraphics[width=8.3cm]{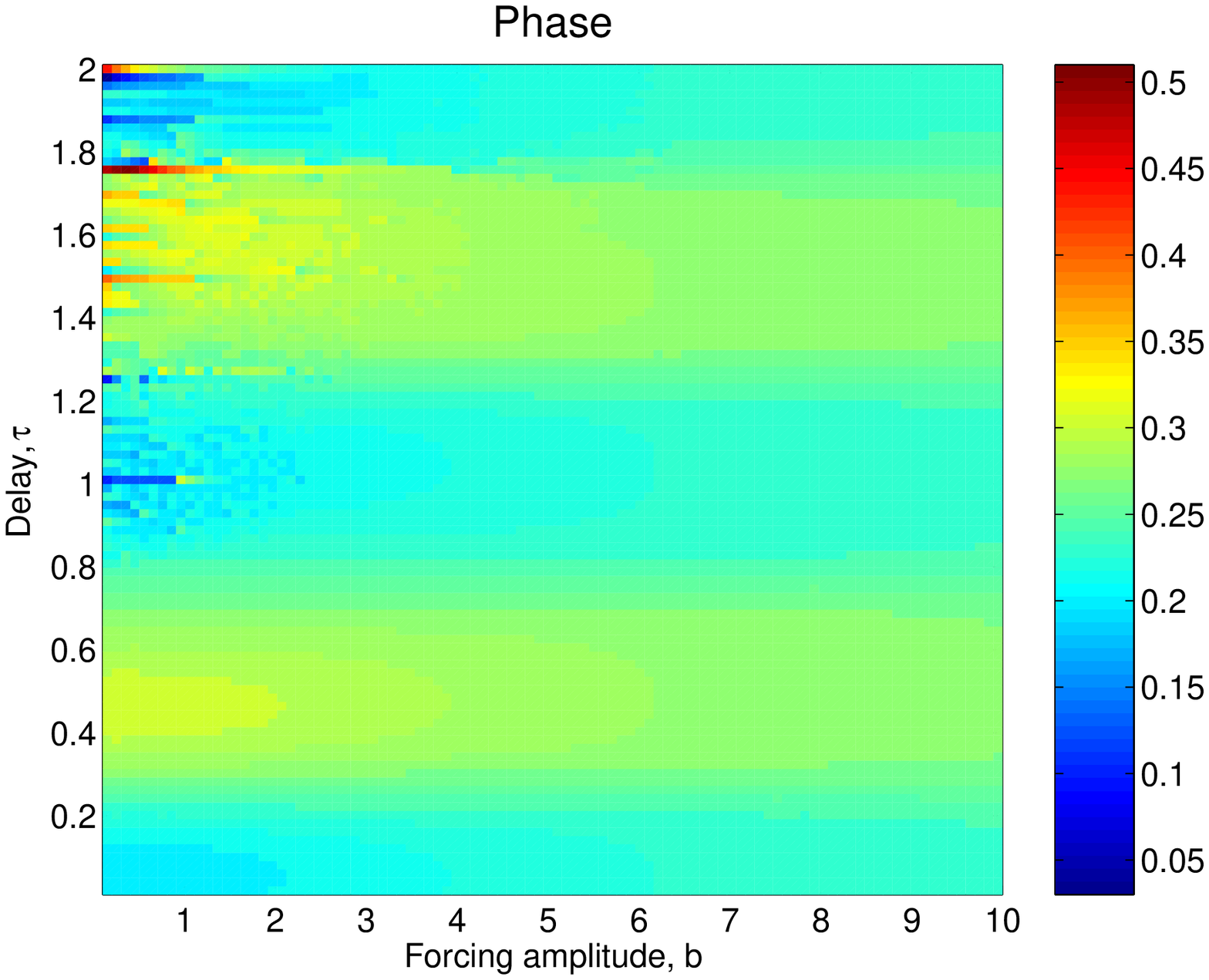}
\centering\includegraphics[width=8.3cm]{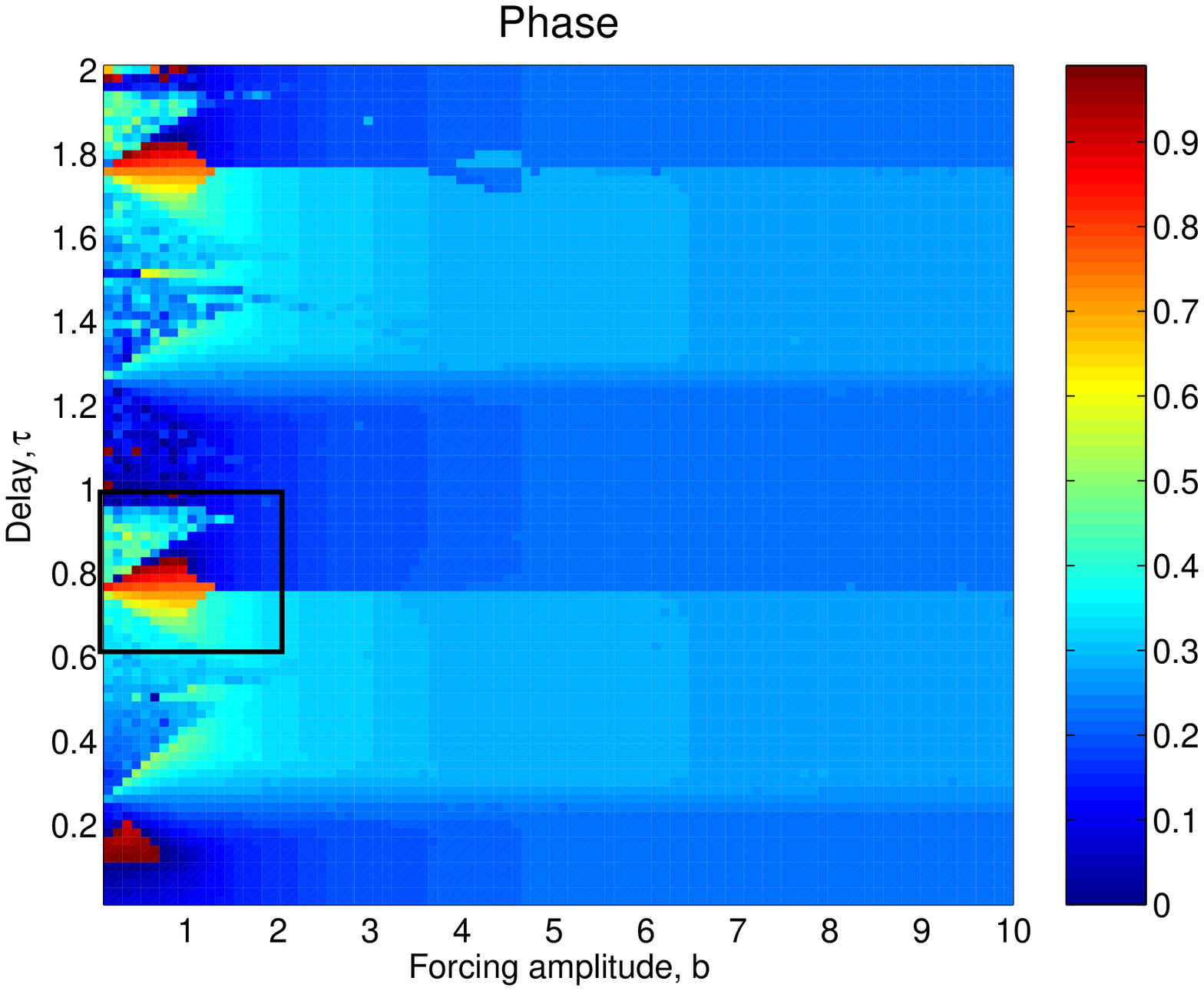}
\caption{Seasonal phase locking of global extrema: parameter dependence.
The plots show the phase $\varphi$ of the global maxima of 
solutions of Eq. ~\eqref{ivp1} for $\kappa=2.0$ (top panel)
and $\kappa=11.0$ (bottom panel); same number of solutions 
and parameter range as in Fig.~\ref{expos_hist}.
The rectangle in the bottom panel 
highlights the region blown up in Fig.~\ref{expos_loc}.} 
\label{expos_map}
\end{figure}

In this section, we study the phase $\varphi$ of the local maxima
and minima of the model solutions that obey \eqref{ivp1}-\eqref{ivp2}. 
The main result, as we shall see, is that the model's extrema occur exclusively 
within a particular season.

We start with several examples that illustrate the analysis in the rest of
the section. Figure~\ref{tr_ex1}$a$ shows a piece of model solution $h(t)$   
for $\tau=0.5$, $\kappa=11$, and $b=2$.
This solution has period $P=1$, as illustrated in panel ($b$),
which shows the scatterplot of the pairs $(h(t_i), h(t_i+1))$ for 
$i=0,1,\dots$ and $t_{i+1}=t_i+1$. Given the 1-periodic character
of the solution, all the points $(h(t_i), h(t_i+1))$ coincide. 
The choice of the starting point $t_0$ will only affect the
position of a single point in the panel (not shown).

For each time epoch $t$ we define its position $\varphi$ within the seasonal cycle
as $\varphi=t({\rm mod}~1)$; the origin of the seasonal cycle in the forcing is taken
in October, when the trade winds are strongest.
Panel ($c$) shows the values of the local maxima (filled circles) and minima (squares)
of $h(t)$ as a function of their position $\varphi$ within the seasonal cycle.
The six other panels in Fig.~\ref{tr_ex1} show the results of a similar analysis for a 
solution with period $P=7$ (panels $d-f$) and an aperiodic one (panels $g-i$).

In all the examples of Fig.~\ref{tr_ex1}, most of the local maxima are located within 
the first half of the annual cycle, {\it i.e.} in boreal winter, while the local minima lie 
within the second half, {\it i.e.} in boreal summer.
Moreover, the global maximum, as well as local maxima with large amplitudes,
are always located within the $\varphi$-interval $(0.15, 0.4)$, while the global
minimum, as well as large-amplitude local minima, are always located
within the interval $(0.7, 0.95)$. 
We found this characteristic property of the model to hold for
most of its solutions.

To verify this model property, we analyzed the positions of the local extrema for a large number 
of individual solutions of Eq.~\eqref{ivp1} within the parameter region 
$(0<\tau\le 2, 0<b\le 10)$ and at several values of $\kappa$.
The representative results are summarized in Figs.~\ref{expos_hist} and \ref{expos_map},
where we used 10 000 individual solutions for each value of $\kappa$.
Figure~\ref{expos_hist} shows histograms of positions of the local extrema
within the seasonal cycle, while
Fig.~\ref{expos_map} plots the position of the global maximum as a function of
the model parameters $\tau$ and $b$. 

\begin{figure}[t]
\vspace*{2mm}
\centering\includegraphics[width=8.3cm]{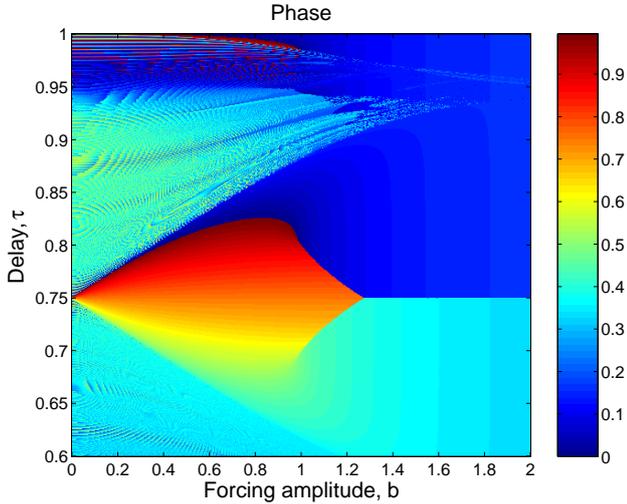}
\caption{Reversal in the phase locking of the maxima.
The plot shows the seasonal cycle position $\varphi$ of the global 
maximum for 250 000 solutions of Eq. ~\eqref{ivp1}, for $\kappa=11.0$;
it represents a blow-up of the region marked by a rectangle 
in the lower panel of Fig.~\ref{expos_map}.} 
\label{expos_loc}
\end{figure}

The phase locking of the extrema to the seasonal cycle is present for most 
combinations of the physically relevant model parameters.
Moreover, the local maxima tend to occur, depending
on the value of $\tau$, at either $\varphi=0.23$ or $\varphi=0.27$, while the 
local minima occur at $\varphi=0.73$ or $\varphi=0.77$.
We notice that the cosine-shaped seasonal forcing vanishes 
at $\varphi=0.25$ and $\varphi=0.75$;
hence the local maxima occur in the vicinity of zero forcing when the latter
decreases, and the local mimina occur in the vicinity of zero forcing when 
the latter increases.
The offset in the position of the extrema from the point where the external 
forcing vanishes seems to be independent of the model parameters.

\begin{figure}[t]
\vspace*{2mm}
\centering\includegraphics[width=8.3cm]{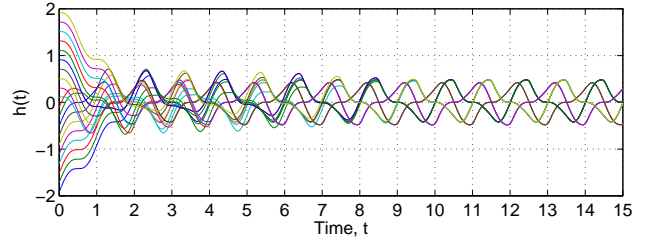}
\caption{Multiple stable solutions. 
Twenty trajectories that correspond to different initial 
histories $\phi(t)\equiv\phi_0$
collapse, after a transient, onto four stable solutions.
Two of these solutions are {\it distinct}, and the other two
can be obtained from the latter by a time shift.
Model parameters are  $\tau=0.5$, $\kappa=10$, and $b=1$;
see also Fig.~\ref{multi_map}.} 
\label{multi_tr}
\end{figure}

As the atmosphere-ocean coupling parameter $\kappa$ increases,
yet another type of sensitive dependence on parameters sets in. 
Namely at low values of the external forcing, $b<1.5$, 
``reversals'' in the location of the local extrema do occur, with maxima
suddenly jumping to boreal summer and minima to boreal winter. 
In Fig.~\ref{expos_loc}, we zoom into one such reversal region, 
marked by a rectangle in Fig.~\ref{expos_map}. 
The dark and light blue colors that occupy most of the region 
indicate that the global maximum of a model solution occurs in the first half of the 
annual cycle, while the red-to-yellow colors that appear around $\tau=0.75$ indicate 
that, within this ``island," the global maximum jumps to the annual cycle's second half.

\begin{figure}[t]
\vspace*{2mm}
\centering\includegraphics[width=8.3cm]{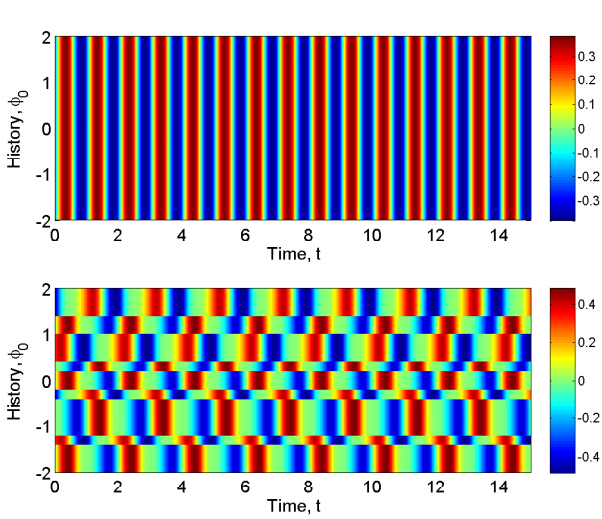}
\caption{ Solution profiles for multiple constant histories 
$\phi(t)=\phi_0$. 
The top panel corresponds to point $A=(\tau=0.4, \kappa=11, b=2)$
in parameter space, where there exists a unique stable solution.
The bottom panel corresponds to
$B=(\tau=0.5, \kappa=10, b=1)$, the same point as in Fig.~\ref{multi_tr}; here
there exist two stable solutions and their attractor basins
are bounded by horizontal discontinuity lines in the solution profile.
The solutions are shown after a sufficiently long transient, 
and the origin of the time is shifted to start from zero; 
color bars indicate
solution values, here as well as in Fig.~\ref{multi_map_1}.}
\label{multi_map}
\end{figure}

\section{Multiple solutions, stable and unstable}
\label{multi}
The analysis in the previous section was carried out, as in Part 1,
for the model \eqref{ivp1}-\eqref{ivp2} with a fixed initial history, 
$\phi(t)\equiv 1$. In this section, we study the model's solutions 
for distinct, yet still constant histories $\phi(t)\equiv\phi_0$.

Naturally, different initial history values $\phi_0$ may result in different 
model solutions. This is illustrated in Fig.~\ref{multi_tr} for the parameter 
values $\tau=0.5$, $\kappa=10$, and $b=1$.
To produce this figure we used 20 distinct initial histories, with constant values 
that are uniformly distributed between $\phi_0=-2$ and $\phi_0=2$; hence, 
at time $t=0$ there exist 20 distinct solutions.
As time passes, those solutions are attracted by a smaller number 
of stable solutions so that, by $t=15$, there are only four distinct
solutions left, all of which have period $P=2$.
We notice furthermore that two of the remaining four solutions can be 
obtained by shifting the other two by one unit of time. 

\begin{figure}[t]
\vspace*{2mm}
\centering\includegraphics[width=8.3cm]{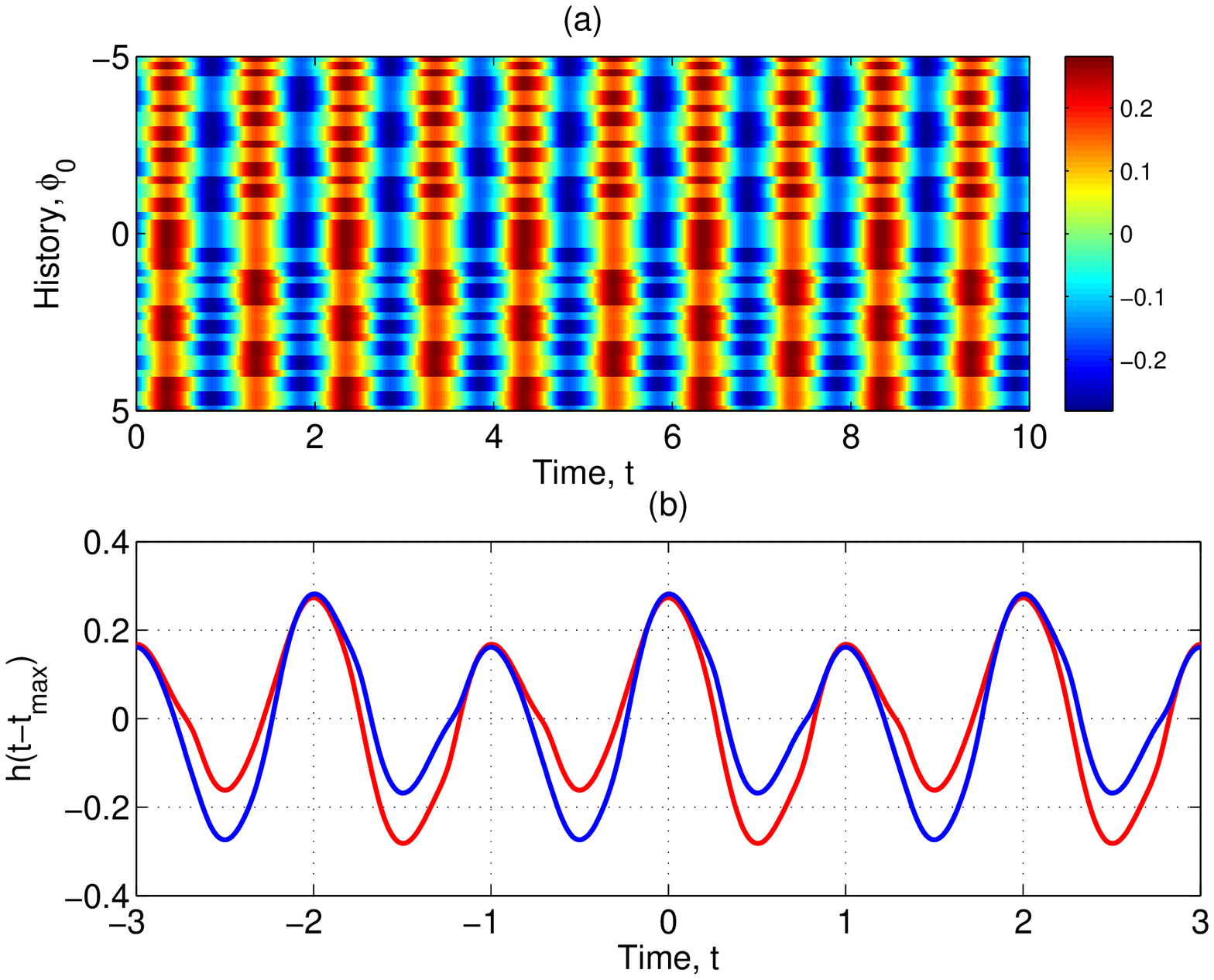}
\centering\includegraphics[width=8.3cm]{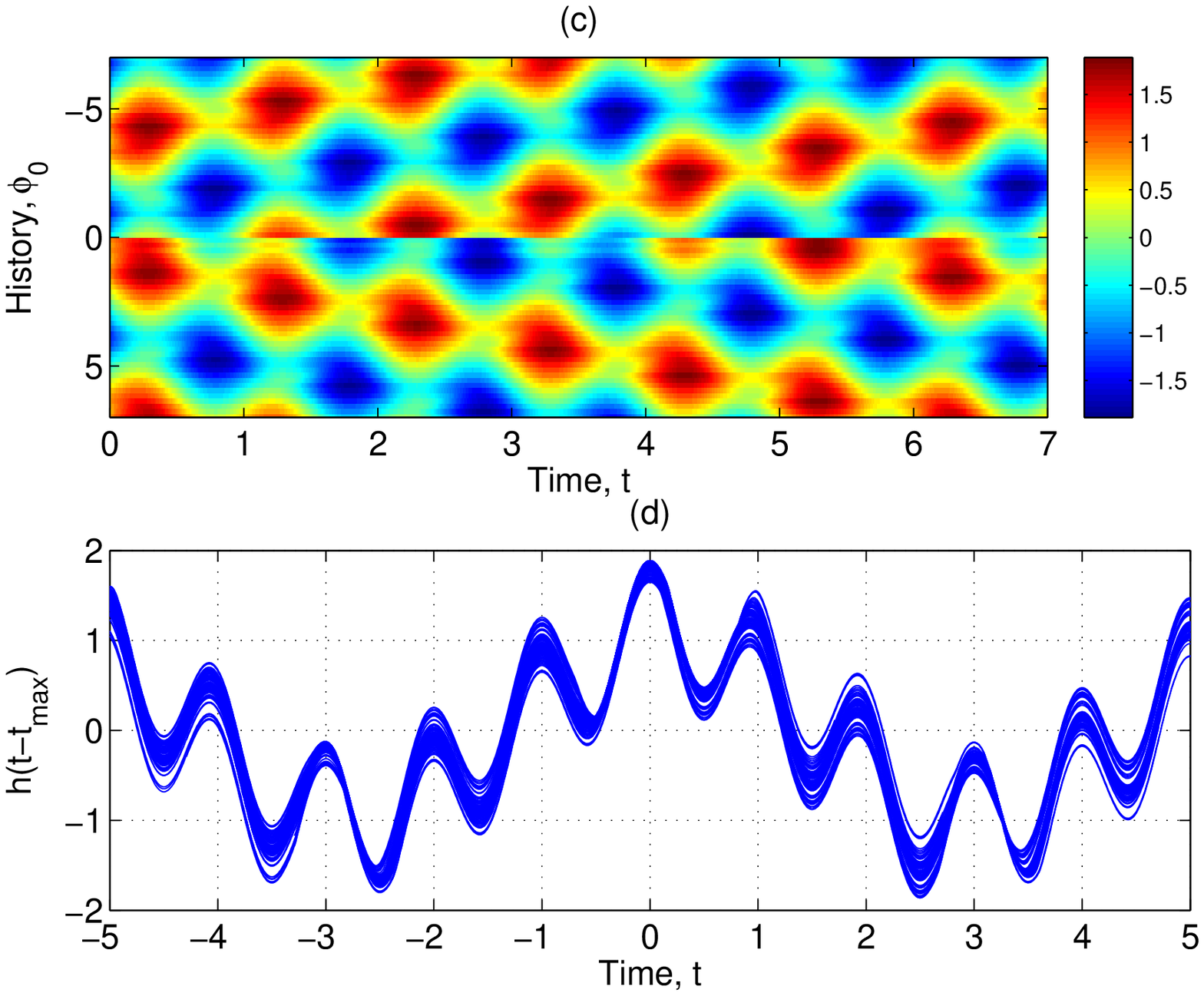}
\caption{Multiple stable solutions.
Solution profile for ($a,c$) different initial histories 
$\phi(t)\equiv\phi_0$, and ($b,d$) the corresponding distinct
solutions.
For visual convenience, the trajectories are shifted to have their 
global maxima at $t=0$.
Panels ($a,b$): model behavior at point $C=(\tau=0.5, \kappa=11, b=1.7842)$,
where there exist 2 distinct solutions; and 
panels ($c,d$): model behavior at point $D=(\tau=1.4579, \kappa=11, b=4)$,
where there exist 61 distinct solutions.}
\label{multi_map_1}
\end{figure}

In general, it is readily seen that --- if the system \eqref{ivp1}-\eqref{ivp2} has 
solution $x(t)$ --- then $x(t+k)$ with any integer $k$ is also a solution.
Hence, if $x(t)$ is a solution with integer period $P=k$, then there are
$k-1$ other solutions obtained from $x(t)$ by an integer time shift.
We will focus on solutions that cannot be obtained from each other 
by such a shift. 
Thus, we call two solutions $x(t)$ and $y(t)$ {\it distinct} if 
$x(t)\not\equiv y(t+k)$ for any positive integer $k\ne P$.

Next we concentrate on the attractor basins of the model's stable solutions.
Figure~\ref{multi_map} shows the model's solution profiles, after a
suitable transient, for $-10\le \phi_0\le 10$, at two points in the model's 
parameter space: point $A=(\tau=0.4, \kappa=1, b=2)$ in the 
top panel, and point $B=(\tau=0.5, \kappa=10, b=1)$ in the bottom panel.
Model behavior at point $B$ was illustrated in Fig.~\ref{multi_tr}. 
At point $A$ the model has a unique stable solution that attracts all
transient solutions as time evolves, so that the solution profile becomes constant 
along any vertical line, {\it i.e.} at any $t=t_0$ in this type of figure.

The model has two distinct stable solutions at point $B$: the boundary 
between their attractor basins, as plotted on the real line of initial-history values $\phi_0$, 
corresponds to points of discontinuity in the solution profiles. These points
line up into straight horizontal lines in Fig.~\ref{multi_map}:
one can see 8 horizontal lines of discontinuity in the solution 
profiles and there would thus appear to be 9 attractor basins. These basins
correspond, however, as shown in Fig.~\ref{multi_tr}, to only two 
stable solutions that are distinct from each other. 

\begin{figure}[t]
\vspace*{2mm}
\centering\includegraphics[width=8.3cm]{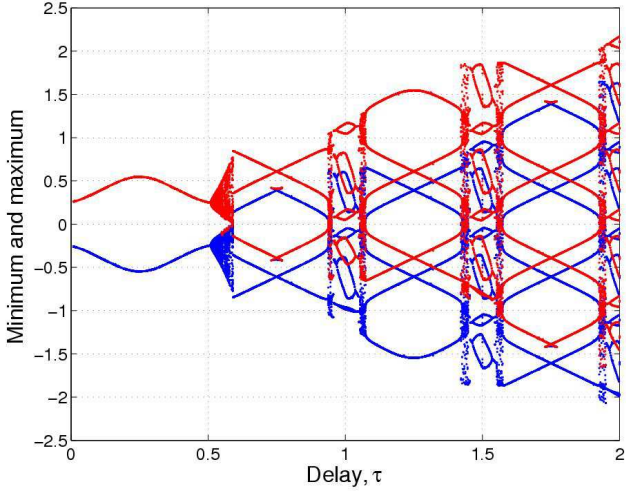}
\caption{Local maxima (red) and minima (blue) of model solutions
as a function of delay $\tau$; the other parameter values are fixed at 
$\kappa=11$ and $b=2$.
Notice the aperiodic regimes between periodic windows
of gradually increasing period. } 
\label{minmax}
\end{figure}

Recall from Sect.~\ref{Banach} that our solutions lie in the 
infinite-dimensional Banach space $X=C([-\tau,0),\mathbb{R})$, and that the
solutions with constant initial histories do not span this space. By using
such a particularly simple type of initial histories, we are merely exploring 
a 1-D manifold of solutions, parametrized by the scalar 
$\phi_0$, in the full space $X$. The intersection of the boundary between the
attractor basins of the two stable solutions with this 1-D manifold gives the
8 lines of discontinuity seen in the bottom panel of Fig.~\ref{multi_map}.

Proposition 1 also implies that a discontinuity in the solution
profile at $\phi_0$ suggests that there exists an unstable solution starting
from $\phi(t)\equiv\phi_0$. Hence, the boundary that separates the two
attractor basins from each other is formed by unstable model solutions.
This boundary is a manifold of codimension one in $X$, and
Figure~\ref{multi_map} reveals 
merely the intersection of this manifold with the 1-D manifold of solutions 
that have constant initial histories. 
The presence of 8 such intersections suggests, in turn, that the boundary
between the two attractor basins is a highly curved, but still smooth 
manifold. 
It is known for finite-dimensional problems that such boundaries can become 
quite complex and possibly fractal \citep{Greb+87}.

Figure~\ref{multi_map_1} shows two slightly more complex situations along
the same lines, namely one with still only two distinct solutions, having both
period $P=2$, but a more intricate pattern of solution profiles (panels $a,b$),
and one with 61 distinct solutions, having all $P=10$ (panels $c,d$).
For visual convenience we shift all the solutions so that their global 
maxima occur at $t=0$.

\begin{figure}[t]
\vspace*{2mm}
\centering\includegraphics[width=8.3cm]{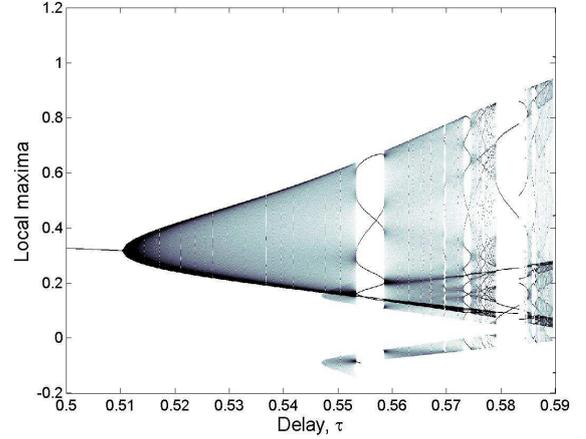}
\caption{Distribution of local maxima as a function of delay $\tau$ within
the interval $0.5<\tau<0.59$; the other parameters are as in Fig.~\ref{minmax}.} 
\label{max_chaos}
\end{figure}
\section{Dynamics of local extrema}
\label{discrete}

We focus here on the dynamics of the local extrema in the model solutions.   
For each solution $h(t)$ we consider the sequence of its local extrema 
$\{e_i\}:=\{h(t_i)$, $i=1,2,\dots\}$, where $h'(t_i)=0$.
The local maxima $\{M_i$, $i=1,2,\dots\}$ are characterized by the additional 
condition that $h''(t_i)<0$, while at the 
local minima $\{m_i$, $i=1,2,\dots\}$ one has $h''(t_i)>0$.

Figure~\ref{minmax} shows the position of the local extrema as a function of
delay $0<\tau<2$ for fixed $\kappa=11$ and $b=2$.
The figure illustrates convincingly the increase in complexity of model solutions 
as the delay $\tau$ increases.
For small delay values, $0<\tau<0.5$, each solution is a periodic sine-like wave
with period $P=1$, which contains a single maximum and a single mimimum 
within each cycle.

Within the interval $0.6 <\tau <0.8$ the solutions become more complex: 
the solution period here is $P=3$, and each cycle has
three local maxima and three local minima.
In general, the time elapsed between a local maximum and the next is an
integer number; this effect is caused by the seasonal forcing, and the
same is true for local minima.
Often, the recurrence interval for extrema of the same kind is just unity and 
the number of local maxima (or minima) coincides with the period $P$ of
a given solution. 

The period in Fig.~\ref{minmax} increases by jumps of 2, from $P=1$ to 
$P=3$ and so on, as $P=2k+1$.
The transitions from one odd-periodic dynamics to the next are associated 
each time with a region of aperiodic behavior; e.g. the one from $P=1$ to $P=3$
occurs in the interval $0.51 < \tau < 0.59$.
Thus, as $\tau$ increases, the number of local extrema becomes larger and each
increase in the number of extrema is preceded by a region of aperiodic,
presumably chaotic behavior.

Figure~\ref{max_chaos} zooms in on the distribution of local maxima within the 
first aperiodic region of Fig.~\ref{minmax}, namely $0.51 < \tau < 0.59$.
In this region, the $\tau$-intervals of aperiodic behavior alternate with shorter 
periodic windows: in the former the local maxima are distributed continuously 
within an interval, while in the latter several distinct local maxima occur within
a comparable range of values.  
This distribution of the maxima resembles the behavior of chaotic dynamical 
systems in discrete time --- {\it e.g.}, period doubling for smooth maps \citep{Feigen,
Kadan} --- and suggests that the model's aperiodic dynamics is in fact chaotic.
An even richer behavior --- with multiple, overlapping cascades --- seems to emerge
for $0.545<\tau$.

\section{Concluding remarks}
\label{discussion}
In the present paper we continued our study of a periodically forced delay differential 
equation (DDE) introduced by Ghil {\it et al.} (\citeyear{GZT08}); 
the DDE~(\ref{dde}) serves as a toy model for ENSO variability.
We studied the model solutions numerically in a broad 3-D
domain of physically relevant parameters: oceanic wave delay $\tau$,
ocean-atmosphere coupling strength $\kappa$, and seasonal forcing amplitude $b$.
In this Part 2 of our investigation, we focussed on multiple model solutions as a 
function of initial histories, and on the dynamics of local extrema.

We found that the system is characterized by {\it phase locking}
of the solutions' local extrema to the seasonal cycle (Figs.~\ref{tr_ex1} and 
~\ref{expos_hist}): solution 
maxima --- {\it i.e.}, warm events (El Ni\~nos) --- tend to occur in boreal winter, 
while local minima --- {\it i.e.}, cold events (La Ni\~nas) --- tend 
to occur in boreal summer. The former model feature is realistic, since
observed warm events do occur by-and-large in boreal winter; in fact,
this property is one of the main features of the observed El Ni\~no events,
having even given rise to the name of the phenomenon
\citep{Phil90,Glantz+91,Diaz92}.

The phase locking of cold events in the model to boreal summer is not
realistic, since La Ni\~nas also tend to occur in boreal winter, rather than in 
phase opposition to the warm ones; see again Fig.~\ref{fig_nino_hist}.
It is not clear at this point which one of the lacking features of our DDE model 
gives rise to this unrealistic phase opposition; it might be the lack of a positive
feedback mechanism, present with a separate, distinct delay in the \cite{Tzi+94}
model. On the other hand, even GCMs with many more detailed features 
may have their warm events in entirely the wrong season; see \cite{GR00}
for a review. 
 
At the same time, for small-to-intermediate seasonal forcing $b$, the
position of the global maxima and minima 
depends sensitively on other parameter values: it may exhibit significant 
jumps in response to vanishingly small changes in the parameter values
(Fig.~\ref{expos_map}). In particular, an interesting phenomenon of
``phase reversal" of the global extrema may occur, {\it cf.} Fig.~\ref{expos_loc}.

We explored a 1-D manifold of solutions for a set of given, prescribed points
$P=(\tau,\kappa,b)$ in the model's parameter space. Such a manifold was
generated, for each $P$, by solutions with constant initial histories $\phi (t)\equiv \phi_0$.

We found multiple solutions coexisting for physically relevant values of the 
model parameters; see Figs.~\ref{multi_tr}--\ref{multi_map_1}. Some of
these solutions are generated by shifting a single solution in time, using 
integer multiples of the period of the forcing, taken here to be unity. We have
often found a set of $k$ solutions so obtained from a single solution of
period $P=k$.

Typically, each stable solution has a bounded, but infinite-dimensional 
attractor basin in the solution space $X$ described in Sect. \ref{Banach}. This 
attractor basin is separated from that of another stable solution by a manifold of 
codimension one, which is generated by unstable solutions (see Proposition 1 
and the following remarks). The intersections of such a manifold with the 1-D
manifold of solutions explored herein appear as the straight horizontal lines
in the solution-profile panels of Figs.~\ref{multi_map} and \ref{multi_map_1}.

In Part 1, we found that the solution period generally increases with the oceanic 
wave delay $\tau$. Figures~\ref{minmax} and \ref{max_chaos} here provide
much more detailed information: the period $P$ of model solutions 
increases in discrete jumps, like $\{P=2k+1, k=0,1,2,\dots\,\}$, separated by 
narrow, apparently chaotic ``windows" in $\tau$. This increase in $P$ 
is associated with the increase of the number of distinct local extrema, all
of which tend to occur at the same position within the seasonal cycle.
The distribution of the maxima in Fig.~\ref{max_chaos} resembles in fact 
the behavior of chaotic dynamical systems in discrete time \citep{Feigen,
Kadan} and suggests that the model's aperiodic dynamics is in fact chaotic.  

It is quite interesting that, for plausible values of the delay $\tau$, the periods
lie roughly between 2 and 7 years, a range that is commonly associated with
the recurrence of relatively strong warm events \citep{Phil90,Glantz+91,Diaz92,
Neel+98}. The sensitive dependence of the period on the model's external 
parameters $(\tau,\kappa,b)$ is consistent with the irregularity of occurrence
of strong El Ni\~nos, and can help explain the difficulty in predicting them 
\citep{Latif94, Ghil+98}. 
\bigskip

{\bf Acknowledgements}
We are grateful to our colleagues M.~D. Chekroun, J.~C. McWilliams, 
J.~D. Neelin and E. Simonnet for many useful discussions; MDC in 
particular suggested the search for multiple solutions, and JDN 
provided information on the seasonal cycle in the surface winds. 
This study was supported by DOE Grants DE-FG02-07ER64439 
and DE-FG02-07ER64440 from the Climate Change Prediction Program, 
and by the European Commission's No. 12975 (NEST) project 
``Extreme Events: Causes and Consequences (E2-C2).''

\end{document}